\documentclass[twocolumn, prl, superscriptaddress]{revtex4-1}
\usepackage{amssymb}
\usepackage{amsmath}
\usepackage{epsfig}
\usepackage{color}
\usepackage{graphics, graphicx}
\usepackage{bbold}
\usepackage{psfrag}
\usepackage{mathcomp}
\usepackage{subfigure}
\usepackage{verbatim}
\usepackage{color}
\usepackage[colorlinks,citecolor=blue]{hyperref}

\begin{document}
\title{Ultrastable super-Tonks-Girardeau gases under weak dipolar interactions}
\author{Yu Chen}
\affiliation{Beijing National Laboratory for Condensed Matter Physics, Institute of Physics, Chinese Academy of Sciences, Beijing 100190, China}
\affiliation{School of Physical Sciences, University of Chinese Academy of Sciences, Beijing 100049, China}
\author{Xiaoling Cui}
\email{xlcui@iphy.ac.cn}
\affiliation{Beijing National Laboratory for Condensed Matter Physics, Institute of Physics, Chinese Academy of Sciences, Beijing 100190, China}
\date{\today}

\begin{abstract}

The highly excited super-Tonks-Girardeau (sTG) gas  was recently observed to be extremely stable in the presence of a weak dipolar repulsion. Here we reveal the underlying  reason for this mysterious phenomenon. 
By exactly solving the trapped small clusters with both contact and dipolar interactions, we show that the reason lies in the distinct spectral responses between sTG gas and its decaying channel (bound state) when turn on a weak dipolar interaction. Specifically, a tiny dipolar force can produce a visible energy shift for the localized bound state, but can hardly affect the extended sTG branch. As a result, the avoided level crossing between two branches is greatly modified in both location and width in the parameter axis of coupling strength, leading to a more (less) stable sTG gas for a repulsive (attractive) dipolar force. These results, consistent with experimental observations, are found to robustly apply to both bosonic and fermionic systems.  
\end{abstract}

\maketitle

Super-Tonks-Girardeau (sTG) state stays in a highly excited branch in one dimension  (1D) under inter-particle attractions, which hosts an even stronger correlation than the Tonks-Girardeau (TG) regime with hard-core repulsions. Such an intriguing state was first predicted in identical bosons by quantum Monte-Carlo\cite{QMC} and Bethe-ansatz methods\cite{BA}, and subsequently realized in a quasi-1D ultracold Bose gas as tuning the 1D coupling strength across resonance\cite{sTG_Science2009}. Later, the sTG state of spin-1/2 fermions was also discovered with Bethe-ansatz solutions\cite{Chen} and observed experimentally in trapped small clusters\cite{sTG_fermion_expt1,sTG_fermion_expt2}. Recently the fermionic sTG gas has attracted great interests in exploring the itinerant Ferromagnetism in 1D and various spin chain configurations without lattice\cite{Blume, Conduit, Cui1, Zinner2, Zinner, Santos, Pu, Cui2, Parish}. 

However, the sTG gas is not always stable in practice --- as moving away from resonance,  the gas will eventually collapse to low-lying bound states at intermediate attraction strength
\cite{sTG_Science2009}, making it impossible to approach  stronger correlations with higher repulsive energies. Surprisingly, such instability has recently been rescued in experiment just by adding a weak dipolar repulsion among the atoms\cite{sTG_dipole_expt}. 
There, the gaseous repulsive branch was shown to be extremely stable over the whole sTG regime, and can even 
evolve adiabatically for two rounds of interaction cycles with continuously increasing energies, realizing  the quantum holonomy ever in a physical system\cite{Cheon}. On the other hand, when switching to a weak dipolar attraction, the sTG gas was found to be less stable instead. 
These observations raise two big puzzles. First, how could a weak dipolar force, which barely changes the energy of sTG gas,  influence its stability so significantly? Secondly, why does this influence depend on the sign of dipolar force? Up to date no definitive answers arise to these puzzles.

\begin{figure}[t]
\includegraphics[width=8cm]{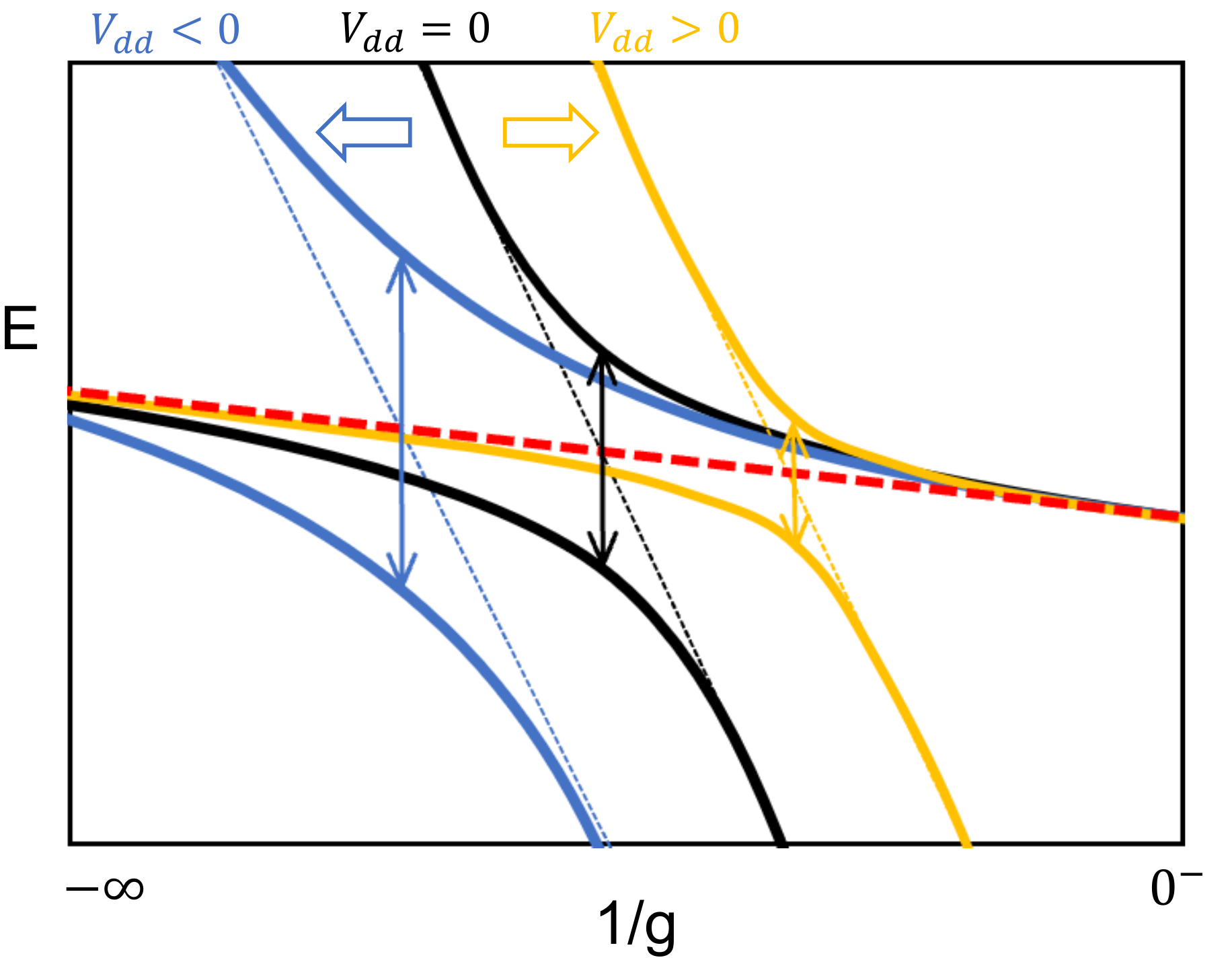}
\caption{(Color online) Illustration for the modified stability of sTG gas by dipolar interaction $V_{dd}$. Red dashed line marks the energy level of sTG gas, which can  become unstable due to the hybridization with an excited bound state (EBS, dotted line) at their avoided level crossing. In the presence of a weak $V_{dd}$, sTG is hardly affected in energy while the EBS spectrum can be shifted visibly due to its localized wave function and large response to $V_{dd}(\sim 1/r^3)$. For a repulsive $V_{dd}(>0)$, the EBS energy is up-shifted and the avoided crossing moves to $1/g\rightarrow 0^-$ with a narrower width,  giving rise to a more stable sTG gas with weaker hybridization with EBS. In comparison, an attractive $V_{dd}(<0)$ leads to a down-shifted EBS level and thus the avoided crossing moves to $1/g\rightarrow -\infty$ with a broader width (stronger hybridization), giving a less stable sTG gas.
} \label{fig_schematic}
\end{figure}

In this work, we attempt to resolve these puzzles by exactly solving three trapped atoms (bosons or spin-1/2 fermions) with both contact and dipolar interactions. 
Such a three-body system  comprises the minimal yet fundamental model to describe the instability of sTG branch, as manifested by its avoided level crossings with many excited bound states when tuning the coupling strength.  
Based on this, we show that the modified stability of sTG gas 
originates from its distinct spectral response to a weak dipolar interaction, as compared with all bound state channels  it decays into. 
Specifically, given the form of dipolar interaction as $V_{dd}\sim 1/r^3$ ($r$ is the inter-particle distance), it can accumulate much more interaction energy for the localized bound states than the extended sTG gas. As a result, as illustrated in Fig.~\ref{fig_schematic}, the avoided level crossing between the two branches would shift to strong attraction regime if $V_{dd}$ is repulsive, leading to a smaller wave function overlap and thus a narrower width at their crossing. This enhances the stability of sTG gas. Alternatively, when switching to an attractive $V_{dd}$, the inter-branch crossing moves to weak attraction side with a broader width, giving a less stable sTG gas. These effects, consistent with  experimental observations\cite{sTG_dipole_expt}, are universally applicable to identical bosons and spin-1/2 fermions. Our  results suggest a powerful tool in general to tune the stability of target state by  artificially manipulating its decay channels.

We consider the following Hamiltonian ($\hbar=1$):
\begin{eqnarray}
H&=&\sum_i \left( -\frac{1}{2m} \frac{\partial^2}{\partial x_i^2} + \frac{1}{2} m\omega^2 x_i^2 \right)\nonumber \\
&&  +  \sum_{\langle i,j\rangle} \Big(g\delta(x_{i}-x_j) +V_{dd}(x_{i}-x_j)\Big); \label{H} 
\end{eqnarray}
here $x_i$ is the 1D coordinate; $\omega$ is the harmonic trap frequency, and the trap length is defined as $l=1/\sqrt{\mu\omega}$ ($\mu=m/2$ is the reduced mass); $g=-1/(\mu a)$ is the contact coupling with 1D scattering length $a$; for the dipolar interaction $V_{dd}(r)$, since its short-range part is greatly modified by higher transverse modes in realistic quasi-1D geometry\cite{Santos2, Schmelcher, Guan}, here we take a short-range cutoff $r_c(=0.15l)$ and simplify it as $D/|r|^3$ for $r>r_c$ and $0$ otherwise. 

The three-body problem of identical bosons or spin-1/2 fermions can be exactly solved based on (\ref{H}). To facilitate later discussions, we shall mainly focus on the fermion case ($\downarrow\uparrow\uparrow$) where  analytical results are available. Consider a spin-$\downarrow$ atom at $x_1$ and two $\uparrow$ atoms at $x_2,x_3$, we define $r=x_2-x_1$ and $\rho=\frac{2}{\sqrt{3}}(x_3-(x_1+x_2)/2)$ to describe the relative motions, respectively, within a $\downarrow$-$\uparrow$ dimer and between the dimer and the rest fermion. Another set of relative coordinates $\{r_+,\ \rho_+\}$ can be accordingly defined by exchanging $x_2\leftrightarrow x_3$. 
We then expand the three-body ansatz in the center-of-mass(CoM) frame as $\Psi(r,\rho)=\sum_{mn}c_{mn} \phi_m(r)\phi_n(\rho)$, where $\phi_m$ and $\phi_n$ are single particle eigen-states along $r$ and $\rho$ with eigen-energies $\epsilon_{k}=(k+1/2)\omega$ ($k=m,n$). Utilizing the Schr{\"o}dinger equation $H\Psi(r,\rho)=E\Psi(r,\rho)$ and  ensuring the fermion antisymmetry $\Psi(r,\rho)=-\Psi(r_+,\rho_+)$, we obtain the following equation for $\{c_{mn}\}$\cite{supple}:
\begin{eqnarray}
	&&(E-\epsilon_m-\epsilon_n)c_{mn}=g\sum_{ij}c_{ij}\phi_i(0)\Big(\phi_m(0)\delta_{j,n} - A_{mn,j}\Big)\nonumber\\
	&&\ \ \ \ \ \ \ \ +D\sum_{ij}c_{ij}\Big(B_{m,i}\delta_{j,n}- B^+_{mn,ij}+B^-_{mn,ij}\Big),\label{fermi_eigen_equation}
	\end{eqnarray}
with $A_{mn,j}=\int d\rho \phi_m(\sqrt{3}\rho/2)\phi_n(-\rho/2)\phi_j(\rho)$; $B_{m,i}=\int_{|r|>r_c} dr \phi_m(r)\phi_i(r)/|r|^3$;  $B^{+}_{mn,ij}=\int_{|r_+|>r_c} drd\rho \phi_m(r)\phi_n(\rho)\phi_i(r_+)\phi_j(\rho_+)/|r_{+}|^{3}$;
$B^{-}_{mn,ij}=\int_{|r_-|>r_c} drd\rho \phi_m(r)\phi_n(\rho)[\phi_i(r)\phi_j(\rho)-\phi_i(r_+)\phi_j(\rho_+)]/(2|r_{-}|^{3})$ (here $r_{\pm}=\pm r/2+\sqrt{3}\rho/2,\ \rho_{\pm}=\pm \sqrt{3}r/2-\rho/2$). 
 Both $E$ and $\{c_{mn}\}$ can be solved from the matrix equation (\ref{fermi_eigen_equation}). 
Similarly, the exact solutions of three identical bosons can also be obtained\cite{supple}. 

\begin{widetext}

\begin{figure}[t]
\includegraphics[width=14.5cm]{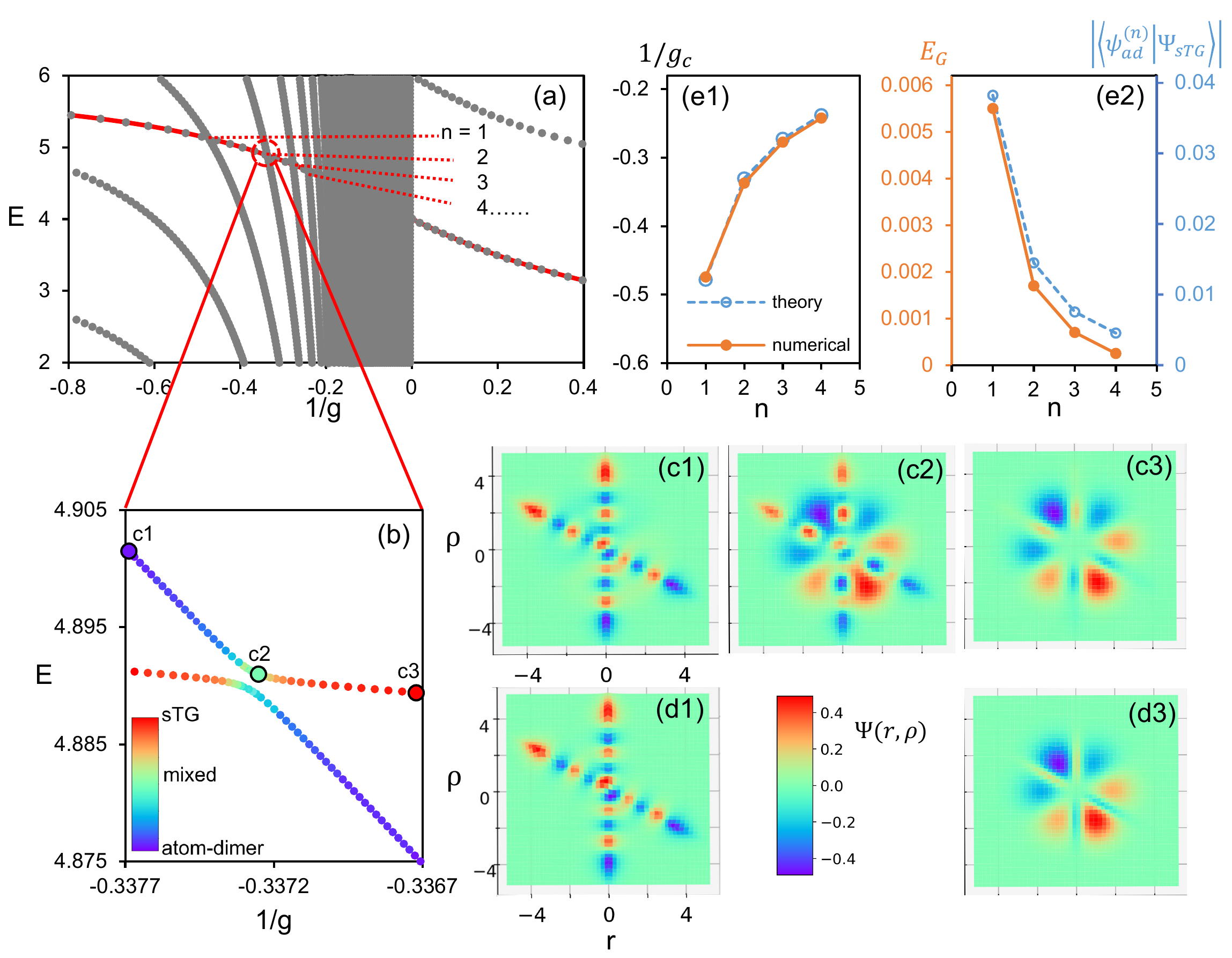}
\caption{(Color online). Hybridization between sTG branch and excited bound states for three harmonically trapped fermions  ($\uparrow\uparrow\downarrow$) without  dipolar interaction. (a) Spectrum in the center-of-mass frame, with the lowest repulsive branch highlighted by red color (the part at $1/g<0$ is the sTG gas).  Indices $'n=1,2,3...'$ mark the locations of avoided level crossing between sTG and various excited atom-dimer states from weak to strong couplings. (b) Magnified spectrum near the second avoided crossing  ($n=2$). The RGB color map is provided according to the wavefunction overlap with sTG (Eq.~(\ref{psi_sTG}), red) and atom-dimer (Eq.~(\ref{psi_ad}), blue) states. (c1-c3) Contour plots of normalized $\Psi(r,\rho)$ for three typical coupling strengths as marked in (b). For comparison, (d1,d3) show theoretical predictions to (c1,c3)  based on Eqs.~(\ref{psi_ad},\ref{psi_sTG}). 
(e1,e2) show the location $1/g_c$ and energy gap $E_G$ for each avoided level crossing. For comparison, the theoretical prediction to $1/g_c$ by comparing (\ref{E_ad}) and (\ref{E_sTG}) is shown in (e1), and    the wavefunction overlap between (\ref{psi_ad}) and (\ref{psi_sTG}) is  shown in (e2).  
In all plots we take $\omega$ and $l$ as the units of energy ($E$) and length ($r,\rho$). The units of $g$ and $\Psi$ are respectively $\omega l$ and $l^{-1}$.} \label{fig_D0}
\end{figure}

\end{widetext}

The trapped three-body system comprises the minimal yet fundamental model to describe the instability of sTG gas. To see this,  let us start from three fermions without dipolar interaction ($D=0$). Its spectrum has been studied previously\cite{Zinner2,Blume, Conduit,Amico,Blume2}, and here we shall focus on the avoided level crossing of sTG branch with a sequence of excited bound states, as labeled by $n=1,2,3...$ in Fig.~\ref{fig_D0}(a) from weak to strong coupling regime.
Near $1/g\rightarrow 0^-$, these bound states are essentially composed of a tight $\uparrow\downarrow$ dimer plus a  free $\uparrow$ atom at excited levels, as described by the atom-dimer wavefunction:
\begin{equation}
\psi^{(m)}_{ad}=\Phi_d(r)\phi_m(\rho) - (x_2\leftrightarrow x_3), \label{psi_ad}
\end{equation}
with energy 
\begin{equation}
E^{(m)}_{ad}=E_{d}+\epsilon_m. \label{E_ad}
\end{equation}
Here $\Phi_d$ is the dimer wavefunction  with energy $E_d$, and $\phi_m$ is the free fermion state with energy $\epsilon_m$. On the other hand, for the repulsive sTG branch near resonance, one can treat $1/g$ as a small parameter and construct an effective spin-chain model $H=1/g \sum_i J_i({\bf s}_i\cdot {\bf s}_{i+1}-1/4)$\cite{Zinner, Santos, Pu, Cui2, Parish}. Here for three atoms  the spin-exchange amplitude $J_i\equiv J$ is site-independent. This gives the wavefunction and energy of sTG gas as:  
 \begin{eqnarray}
\Psi_{sTG}&=&\Psi_0-\frac{1}{g}\Psi_1; \label{psi_sTG}\\
E_{sTG}&=&E_{0}-\frac{3J}{2g},  \label{E_sTG}
 \end{eqnarray}
where $\Psi_0$ is the fermionalized wavefunction in hard-core limit with total energy $E_{0}$, and $\Psi_1$ is from the first order correction when a $\uparrow$-$\downarrow$ pair come close together\cite{supple}. For later comparisons, we have transformed Eqs.~(\ref{psi_sTG}, \ref{E_sTG}) into the CoM frame\cite{supple}.
Fig.~\ref{fig_D0}(b) 
shows that Eqs.~(\ref{psi_ad},\ref{psi_sTG}) can indeed well approximate the two branches far from their level crossings.  

Importantly, Eqs.~(\ref{psi_ad},\ref{psi_sTG}) suggest qualitatively different real-space distributions between  sTG and atom-dimer states. To be concrete, all atom-dimer states have a dominant weight when one $\uparrow$-$\downarrow$ pair come close to each other, i.e., $r\rightarrow 0$ or $r_+=(r+\sqrt{3}\rho)/2\rightarrow 0$, given that they  contain very localized dimer components. In contrast, sTG state is dominated by $\Psi_0$ part which is much more extended in real space, while only has a little weight along the dimer lines ($\sim \Psi_1/g$) . Such difference is numerically confirmed in Fig.~\ref{fig_D0}(c1,c3), where we have plotted real-space  $\Psi$ for different branches and the results are consistent with theoretical predictions from Eqs.~(\ref{psi_ad}, \ref{psi_sTG}) shown in 
Fig.~\ref{fig_D0}(d1,d3).

Above wave-function analyses are crucial for understanding the loss mechanism of sTG gas. 
As shown in Fig.~\ref{fig_D0}(b), at certain $g_c$ when the sTG state and one atom-dimer branch have perfect energy match, they can hybridize strongly and open an energy gap. Accordingly, an avoided level crossing is generated near $g_c$, and the resulted eigenstate   inherits all the key features from both branches (Fig.~\ref{fig_D0}(c2)). Therefore, when driving the sTG gas to  $\sim g_c$, it tends to develop a visible atom-dimer feature and accumulate great possibilities when $\uparrow$-$\downarrow$ come close together. This leads to the instability of sTG gas,  since it can easily undergo an inelastic decay to deep molecule and cause atom loss. Similar inelastic loss due to couplings to excited molecular states was also found  previously for two atoms in anharmonic potentials\cite{inelastic_expt1, inelastic_theory, inelastic_expt2}.

Practically, the loss possibility of sTG gas depends on how strongly it couples to the excited bound states, which can be evaluated by the energy gap $E_G$ at each avoided crossing. Numerically, $E_G$ can be extracted as the minimal energy difference at each avoided crossing,  and accordingly the location $g_c$ can also be identified, as shown in Fig.~\ref{fig_D0}(e1,e2). We can see that as the crossing point moves away from resonance (smaller  $'n'$),  $E_G$ becomes larger, consistent with a larger wavefunction overlap between $\psi_{sTG}$ and $\psi^{(n)}_{ad}$ (see comparison in Fig.~\ref{fig_D0}(e2)). This indicates a less stable sTG gas, since it has a stronger  hybridization with excited bound states in a broader interaction window and thus can easily transit to decay channels.  Such picture is supported by experimental observations that the sTG gas eventually collapses at intermediate $g(<0)$ when moving away from resonance\cite{sTG_Science2009,sTG_dipole_expt}.

Given the loss mechanism of sTG gas as above, now we are ready to study the effect of dipolar interaction $V_{dd}$. In accordance with Ref.\cite{sTG_dipole_expt}, we focus on a weak $V_{dd}$  with $|D|\ll \omega l^3, gl^2$. We will show below that even a weak $V_{dd}$ can dramatically change the stability of sTG gas, and the  key lies in the distinct spectral responses between different branches when $V_{dd}$ is turned on. 

\begin{figure}[t]
\includegraphics[width=9cm]{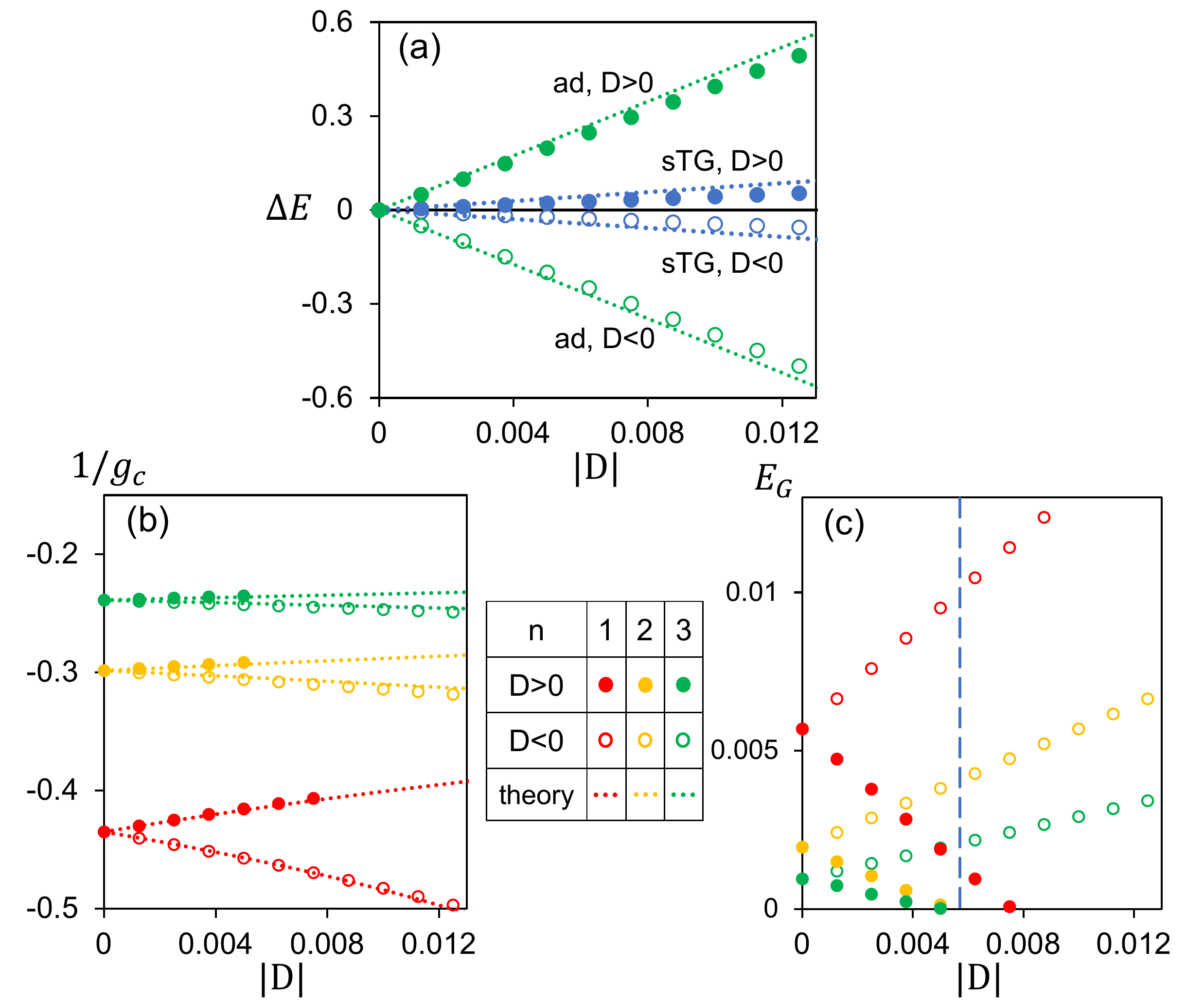}
\caption{(Color online). Response of three fermions to a weak dipolar interaction with strength $D$. (a) Energy shifts of sTG and excited atom-dimer branches as functions of $|D|$ at given $\omega l/g=-0.27$. Dotted lines show mean-field shifts $\langle V_{dd} \rangle$. (b) Locations of three avoided crossings (as marked by $'n=1,2,3'$ in Fig.~\ref{fig_D0}(a)) as functions of $|D|$.  Dotted lines shows linear fits according to Eq.~(\ref{shift}).
(c) Associated energy gap $E_G$ of each avoided crossing as a function of $|D|$. 
Dotted vertical line marks the strength of repulsive  $D$ used in experiment\cite{sTG_dipole_expt}.
Here the energy $\Delta E$,   coupling $g$, and dipolar force $D$ are respectively in units of $\omega$, $\omega l$ and $\omega l^3$.} \label{fig_dipole_fermion}
\end{figure}

Taking a typical $g$ away from any $g_c$, in Fig.~\ref{fig_dipole_fermion}(a) we plot the energy shifts $\Delta E$ for the sTG gas and its nearest atom-dimer branch as varying $|D|$. Clearly, the atom-dimer energy changes rapidly as $|D|$ increases, while the sTG energy changes much more slowly. This can be attributed to very different real-space distributions of the two states. Namely, the atom-dimer  is more localized along $r$, $r_+\rightarrow 0$ and therefore it  produces a significant spectral response to $V_{dd}\sim (1/|r|^3+1/|r_+|^3+...)$; on the contrary, sTG is more extended and has little weight near $r$, $r_+\rightarrow 0$, leading to a negligible energy shift. At small $D$, $\Delta E$ of each branch can be well approximated by mean-field shift $\langle V_{dd}\rangle$, as shown by dotted lines in Fig.~\ref{fig_dipole_fermion}(a). This allows us to analytically determine the shift of crossing point, $\Delta(1/g_c)$, by equating (\ref{E_ad}) and (\ref{E_sTG}) after adding up $\langle V_{dd}\rangle$  for each branch:
\begin{equation}
\Delta(1/g_c)=\frac{\langle V_{dd} \rangle_{ad}-\langle V_{dd} \rangle_{sTG}}{\partial{E_d}/\partial{(1/g)}-3J/2}. \label{shift}
\end{equation}

Eq.~(\ref{shift}) tells that the distinct spectral responses, $\langle V_{dd}\rangle_{sTG}\neq \langle V_{dd}\rangle_{ad}$, directly lead to a finite shift $\Delta(1/g_c)$ of inter-branch crossing. Moreover, the sign of $\Delta(1/g_c)$ exactly follows  $D$: a positive $D$ will drive the crossing point towards  resonance ($\Delta(1/g_c)>0$), while a negative $D$ drives it oppositely  ($\Delta(1/g_c)<0$). All these features are verified numerically in Fig.~\ref{fig_dipole_fermion}(b), where Eq.~(\ref{shift}) provides a reasonably good fit to the shift $\Delta(1/g_c)$ at small $|D|$. In addition, we observe that $\Delta(1/g_c)$ becomes less pronounced for crossings near resonance. This can be attributed to the large denominator of Eq.~(\ref{shift}) produced by $\partial{E_d}/\partial{(1/g)}\propto g^3$ for deep dimers. 
 Therefore, $V_{dd}$ can only visibly affect the level crossings with small $'n'$ but not those with large $'n'$ near resonance. 

Given the intimate relation between $1/g_c$ and $E_G$ (see Fig.~\ref{fig_D0}(e1,e2)), the shift of $1/g_c$  by $V_{dd}$ inevitably leads to the change of $E_G$, as plotted in Fig.~\ref{fig_dipole_fermion}(c). For a repulsive $V_{dd}$($D>0$), all $1/g_c$ shift towards resonance with decreasing $E_G$, indicating a more stable sTG gas; while for an attractive $V_{dd}$($D<0$),  $1/g_c$ shift away from resonance with increasing $E_G$, indicating a less stable sTG. Again, $E_G$ changes most visibly for the outmost crossing ($'n=1'$). Remarkably, all $ E_G$ become vanishingly small at a weak $D/(\omega l^3)\sim 0.008$, suggesting an extremely sTG gas in the whole $g<0$ regime. We note that if take the same $D$ as used in experiment\cite{sTG_dipole_expt}, as marked by vertical dot line in Fig.~\ref{fig_dipole_fermion}(c),  the outmost $E_G$ ($'n=1'$) is greatly reduced compared to  $D=0$ case, and all other $E_G$ decrease to $<10^{-4}\omega$. 
Besides $D$, we have also checked the effect of  dipolar cutoff $r_c$. It is found that  a smaller $r_c$ can generate more energies for the localized atom-dimer branch\cite{supple} and thus the change of $1/g_c$ and $E_G$ will be more dramatic at the same $D$.

In above we have shown how a weak repulsive or attractive $V_{dd}$ can greatly affect the stability of fermionic sTG  gas, while leaving its energy essentially unchanged. 
We have checked that similar physics also holds for three identical bosons\cite{supple}, including the distinct energy responses between sTG and excited bound states as well as the dependence of sTG stability on the sign of $V_{dd}$. Here a crucial difference from fermion case is that the bosonic bound states are all cluster ones that no longer follow the atom-dimer description.  Further, for a much larger bosonic system as in experiment\cite{sTG_dipole_expt}, which is inaccessible by exact numerics, we expect above loss mechanism to equally work.  
In this case, sTG branch can also be well identified as the adiabatic extension of TG gas to $1/g<0$ side, and as changing $1/g$ there should be many more cluster bound states from higher harmonic levels to (avoided) level cross with it.  However, only those far from resonance are responsible for the instability of sTG gas due to their stronger hybridizations in-between. Similar to the few-body case,  a weak dipolar force is expected to shift these bound state spectra visibly due to their localized nature but can hardly affect the energy of sTG branch. Again, this leads to a significant change of sTG stability that is sensitive to the sign of $V_{dd}$ (see Fig.~\ref{fig_schematic}).

In summary, we have revealed the underlying mechanism for a mysterious phenomenon recently observed in 1D sTG gas, namely, its greatly enhanced (reduced) stability by a weak repulsive (or attractive) dipolar interaction. The key to this phenomenon  is the significant spectral response of excited bound states --- the decay channel of sTG gas --- to the dipolar force. Therefore the sTG gas is indirectly affected due to the inter-branch hybridization at their level crossing, leading to a modification of sTG stability but not its energy. In this regard, our results suggest a powerful tool to tune the stability of target state by  manipulating its decay channel under designed potentials. Such state-selective manipulation may help to  engineer many  more  fascinating long-lived quantum states in cold atoms platform in future.

{\bf Acknowledgement.} We thank Benjamin Lev for positive and helpful feedback on our manuscript. This work is supported by the National Natural Science Foundation of China (12074419, 12134015), and the Strategic Priority Research Program of Chinese Academy of Sciences (XDB33000000).

\clearpage

\onecolumngrid
\vspace*{1cm}
\begin{center}
{\large\bfseries Supplementary Materials}
\end{center}
\setcounter{figure}{0}
\setcounter{equation}{0}
\renewcommand{\figurename}{Fig.}
\renewcommand{\thefigure}{S\arabic{figure}}
\renewcommand{\theequation}{S\arabic{equation}}

This supplemental file includes details on the exact solutions of three fermions or bosons with both contact and dipolar interactions, the wavefunction and energy of super-Tonks-Girardeau(sTG) gas near resonance, and the results of three identical bosons.   

\section{I.\ \ \ Three-body problem with contact and dipolar interactions}

\begin{figure}[h]
\includegraphics[width=14cm]{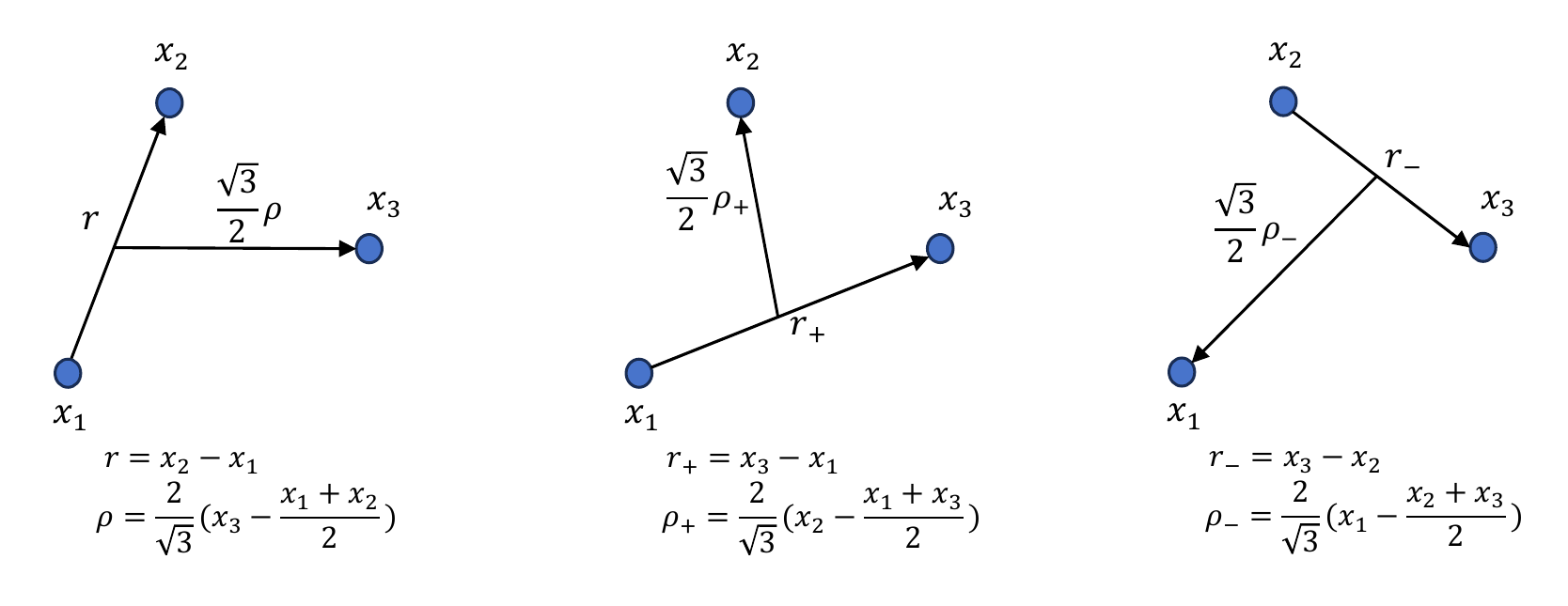}
\caption{(Color online). Relation between relative coordinates $\{(r,\rho), (r_+,\rho_+), (r_-,\rho_-)\}$ and original ones $\{x_1,x_2,x_3\}$. For better illustration, we have arranged all coordinates in a 2D plane instead of a 1D line.} \label{fig_rrho}
\end{figure}

We consider three atoms ($x_1,x_2,x_3$) described by the Hamiltonian $H$ (Eq.~(1) in the main text). Decoupling the center-of-mass (CoM) motion from the problem and defining two relative coordinates $r=x_2-x_1$ and $\rho=\frac{2}{\sqrt{3}}(x_3-(x_1+x_2)/2)$, we can rewrite $H$ as $H=H_0+U$, where
\begin{eqnarray}
H_0&=&-\frac{1}{2\mu} (\partial^2_r+\partial^2_{\rho})+\frac{1}{2}\mu\omega^2(r^2+\rho^2); \label{h0}\\
U&=&g(\delta(r)+\delta(r_+)+\delta(r_-))+V_{dd}(r)+V_{dd}(r_+)+V_{dd}(r_-). \label{U}
\end{eqnarray}
Here $\mu=m/2$, $r_{+}\equiv x_3-x_1=r/2+\sqrt{3}\rho/2$ and $r_{-}\equiv x_3-x_2=-r/2+\sqrt{3}\rho/2$. Accordingly, we define $\rho_{\pm}=\pm\sqrt{3}r/2-\rho/2$. The relation between these relative coordinates $\{(r,\rho), (r_+,\rho_+), (r_-,\rho_-)\}$ and original ones $\{x_1,x_2,x_3\}$ is illustrated in Fig.~\ref{fig_rrho}.
The  1D coupling $g$ and dipolar interaction $V_{dd}(r)$ have been defined in the main text. 

In the CoM frame, the three-body wave function can be expanded in terms of the harmonic eigenstates:
\begin{equation}
	 \Psi(r,\rho)=\sum_{mn} c_{mn}\phi_{m}(r)\phi_{n}(\rho),\ \ \ \ \ {\rm with} \ \ \phi_{n}(x)=\frac{1}{\pi^{\frac{1}{4}}\sqrt{2^nn!l}}e^{-\frac{x^2}{2l^2}}H_n(x/l)
	 \label{function_expansion}
\end{equation}
Here the trap length is defined as $l=1/\sqrt{\mu\omega}$. 

The three-body problem can be exactly solved by employing the Schr{\" o}dinger equation $H\Psi=E\Psi$, which is equivalent to the Lippmann-Schwinger equation 
\begin{equation}
\Psi=G_0 U \Psi, \label{LS}
\end{equation}
with $G_0=(E-H_0)^{-1}$ the non-interacting Green's function. Next we will solve the  three-fermion and three-boson problems separately.

\subsection{(a) Three fermions $\downarrow\uparrow\uparrow$}

We consider three spin-$1/2$ fermions  with one spin-$\downarrow$ at $x_1$ and two spin-$\uparrow$  at $x_2,\ x_3$. The Fermi statistics automatically rules out the contact interaction between two $\uparrow$ at $x_2=x_3$, i.e., no effect of $\delta(r_-)$ term in Eq.~(\ref{U}). Moreover, the Fermi statistics  
requires the anti-symmetry of $\Psi$ when exchanging $x_2\leftrightarrow x_3$, i.e.,
\begin{equation}
\Psi(r,\rho)=-\Psi(r_+,\rho_+). \label{anti_sym}
\end{equation}
Introducing an auxiliary function $f(r,\rho)\equiv U\Psi(r,\rho)$, we can easily see that Eq.~(\ref{anti_sym})  directly leads to the anti-symmetry of $f$-function $f(r,\rho)\equiv U\Psi(r,\rho)$, i.e., $f(r,\rho)=-f(r_+,\rho_+)$. To satisfy this, and recalling Eqs.~(\ref{U},\ref{function_expansion}), we write $f(r,\rho)$ as  
\begin{eqnarray}
f(r,\rho)&=&g\left(\sum_{mn} c_{mn}\phi_{m}(0)\phi_{n}(\rho)\delta(r) - \sum_{mn} c_{mn}\phi_{m}(0)\phi_{n}(\rho_+)\delta(r_+)  \right) + V_{dd}(r) \sum_{mn} c_{mn}\phi_{m}(r)\phi_{n}(\rho)  \nonumber\\
&&\ \ \ - V_{dd}(r_+) \sum_{mn} c_{mn}\phi_{m}(r_+)\phi_{n}(\rho_+) + V_{dd}(r_-) \left( \sum_{mn} c_{mn}\phi_{m}(r)\phi_{n}(\rho) -\sum_{mn} c_{mn}\phi_{m}(r_+)\phi_{n}(\rho_+) \right)/2. \label{f}
\end{eqnarray}
Note that for the last term in above equation, $V_{dd}(r_-)$ depends on $|r_-|$ and thus is symmetric under the exchange  $x_2\leftrightarrow x_3$.

The Lippmann-Schwinger equation (\ref{LS}) can be written as
\begin{equation}
\Psi(r,\rho)=\int dr' d\rho' \langle r,\rho |G_0| r',\rho' \rangle f(r',\rho').  \label{psi_2}
\end{equation}
Plugging Eqs.~(\ref{function_expansion}, \ref{f}) into above equation, we obtain the self-consistent equation for $\{ c_{mn}\}$:
\begin{eqnarray}
	&&(E-\epsilon_m-\epsilon_n)c_{mn}=g\sum_{ij}c_{ij}\phi_i(0)\Big(\phi_m(0)\delta_{j,n} - A^{(1)}_{mn,j}\Big) +D\sum_{ij}c_{ij}\Big(B_{m,i}\delta_{j,n}- B^{+\ (1)}_{mn,ij}+B^{-\ (1)}_{mn,ij}\Big),  \label{fermi_eigen_equation}
	\end{eqnarray}
where 
\begin{eqnarray}
A^{(1)}_{mn,j}&=&\int d\rho \phi_m(\sqrt{3}\rho/2)\phi_n(-\rho/2)\phi_j(\rho); \\
B_{m,i}&=&\int_{|r|>r_c} dr \phi_m(r)\phi_i(r)/|r|^3;\\
B^{+\ (1)}_{mn,ij}&=&\int_{|r_+|>r_c} drd\rho \phi_m(r)\phi_n(\rho)\phi_i(r_+)\phi_j(\rho_+)/|r_{+}|^{3};\\
B^{-\ (1)}_{mn,ij}&=&\int_{|r_-|>r_c} drd\rho \phi_m(r)\phi_n(\rho)[\phi_i(r)\phi_j(\rho)-\phi_i(r_+)\phi_j(\rho_+)]/(2|r_{-}|^{3}).
\end{eqnarray}
Solving the large matrix equation (\ref{fermi_eigen_equation}), we can obtain both  $E$ and $\{c_{mn}\}$, and the wavefunction $\Psi(r,\rho)$ can be straightforwardly obtained via Eq.~(\ref{function_expansion}).

Two remarks on $\Psi(r,\rho)$ are as below.  First,  its anti-symmetry (\ref{anti_sym}) can be guaranteed given $f(r,\rho)=-f(r_+,\rho_+)$. To prove this, we start from Eq.~(\ref{psi_2}):
\begin{eqnarray}
\Psi(r,\rho)&=&\int dr' d\rho' \langle r,\rho |G_0| r',\rho' \rangle f(r',\rho') \nonumber\\
&=&-\int dr' d\rho' \langle r,\rho |G_0| r',\rho' \rangle f(r_+',\rho_+') \nonumber\\
&=&-\int dr' d\rho' \langle r,\rho |P_{23}G_0P_{23}| r',\rho' \rangle f(r_+',\rho_+') \nonumber\\
&=&-\int dr_+' d\rho_+' \langle r_+,\rho_+ |G_0| r_+',\rho_+' \rangle f(r_+',\rho_+') \nonumber\\
&=&-\Psi(r_+,\rho_+). 
\end{eqnarray}
Here the operator $P_{23}$ is to exchange coordinates $x_2\leftrightarrow x_3$ (and therefore $r\leftrightarrow r_+,\ \rho\leftrightarrow \rho_+$).

Secondly, the resulted $\Psi(r,\rho)$ satisfies boundary condition:
\begin{equation}
\frac{\partial{\Psi(r,\rho)}}{\partial r}|_{r=0^-}^{r=0^+} = mg \Psi(r=0,\rho). \label{BC}
\end{equation}
To see this, we plug the expression of $c_{mn}$ (Eq.~(\ref{fermi_eigen_equation})) into $\Psi(r,\rho)$ (Eq.~(\ref{function_expansion})), and arrive at
\begin{eqnarray}
\Psi(r,\rho)&=&g\sum_n d_n \Phi_d(E-\epsilon_n,r)\phi_n(\rho)-g\sum_{mn}\sum_j d_j A^{(1)}_{mn,j}\frac{\phi_m(r)\phi_n(\rho)}{E-\epsilon_m-\epsilon_n}  \nonumber\\
&&+D\sum_{mn} \frac{\phi_m(r)\phi_n(\rho)}{E-\epsilon_m-\epsilon_n} \sum_{ij} c_{ij}\Big(B_{m,i}\delta_{j,n}- B^{+\ (1)}_{mn,ij}+B^{-\ (1)}_{mn,ij}\Big), \label{psi3}
\end{eqnarray}
with 
\begin{eqnarray}
d_n&=&\sum_{m}c_{mn}\phi_m(0); \label{dn}\\
\Phi_d(E-\epsilon_n,r)&=&\sum_{m}  \frac{\phi_m(0)\phi_m(r)}{E-\epsilon_n-\epsilon_m}. \label{phi_d}
\end{eqnarray}
One can see from Eq.~(\ref{psi3}) that only the first term has singularity at $r\rightarrow 0$, while the rest terms are all smooth functions around $r\sim 0$. The singularity comes from the dimer function $\Phi_d(E',r)$\cite{Amico, Blume}:
\begin{equation}
\Phi_d(E',r)=\frac{1}{\omega l}\frac{-\Gamma(-v)}{2\sqrt{\pi}}e^{-\frac{r^2}{2l^2}}U(-v,\frac{1}{2},\frac{r^2}{l^2}), \label{exact}
\end{equation}
here $v=E'/(2\omega)-1/4$, $\Gamma(x)$ is the Gamma function and $U(a,b,z)$ the confluent hypergeometric function. It shows a kink at  $r\sim 0$ with discontinuous derivative:
\begin{equation}
\frac{\partial \Phi_d(E',r)}{\partial r}|_{r=0^-}^{r=0^+} = m. \label{partial}
\end{equation}
Then based on Eq.~(\ref{psi3}) we have
\begin{eqnarray}
\frac{\partial \Psi(r,\rho)}{\partial r}|_{r=0^-}^{r=0^+} = mg\sum_n d_n \phi_n(\rho) = mg \sum_{mn}c_{mn}\phi_m(0) \phi_n(\rho) = mg \Psi(r=0,\rho).
\end{eqnarray}
Therefore the boundary condition in Eq.~(\ref{BC}) is proved. Note that here the boundary condition is the same as that without dipolar interaction. This is because here $V_{dd}(r)$ vanishes at $r\rightarrow 0$, and thus its resulted wavefunction (second line in Eq.~(\ref{psi3}) will not cause any singularity at $r\rightarrow 0$.

 It is noted that in our numerical simulation with a finite cutoff of $m$ ($m_{cut}$), the actual dimer function $\Phi_d(E',r)$  in Eq.~(\ref{phi_d}) will deviate from its exact form in Eq.~(\ref{exact}). The deviation is mainly within the short-range regime $|r|<r_s$, where $\Phi_d(E',r)$ shows as a smooth function  across $r\sim 0$ rather than a sharp kink. The range of the deviation ($r_s$) will become narrower for larger $m_{cut}$, and in the limit of $m_{cut}\rightarrow \infty$   one has $r_s\rightarrow 0$ and Eqs.~(\ref{exact},\ref{partial}) will be recovered. The numerical error  of energy solution due to finite cutoffs in our calculation will be discussed in section I(c).


\subsection{(b) Three identical bosons}


For three identical bosons, the Bose  statistics  
requires a symmetric  $\Psi$ when exchanging the coordinates of three atoms,  i.e.,
\begin{equation}
\Psi(r,\rho)=\Psi(r_+,\rho_+)=\Psi(r_-,\rho_-). \label{sym}
\end{equation}
Similar to the fermion case, we introduce an  auxiliary function $f(r,\rho)\equiv U\Psi(r,\rho)$, which should also  be symmetric under particle exchange:  $f(r,\rho)=f(r_+,\rho_+)=f(r_-,\rho_-)$. To satisfy this, we write $f(r,\rho)$ as  
\begin{eqnarray}
f(r,\rho)&=&g\left(\sum_{mn} c_{mn}\phi_{m}(0)\phi_{n}(\rho)\delta(r) + \sum_{mn} c_{mn}\phi_{m}(0)\phi_{n}(\rho_+)\delta(r_+) + \sum_{mn} c_{mn}\phi_{m}(0)\phi_{n}(\rho_-)\delta(r_-)  \right)   \nonumber\\
&&\ \ \ + V_{dd}(r) \sum_{mn} c_{mn}\phi_{m}(r)\phi_{n}(\rho) + V_{dd}(r_+) \sum_{mn} c_{mn}\phi_{m}(r_+)\phi_{n}(\rho_+) + V_{dd}(r_-) \sum_{mn} c_{mn}\phi_{m}(r_-)\phi_{n}(\rho_-). \label{f_b}
\end{eqnarray}
Here to ensure the full symmetry, the quantum number $m$ should be all even. Plugging Eqs.~(\ref{function_expansion}, \ref{f_b}) into the Lippmann-Schwinger equation (\ref{LS}), we obtain self-consistent equation for $\{ c_{mn}\}$: 
\begin{eqnarray}
	&&(E-\epsilon_m-\epsilon_n)c_{mn}=g\sum_{ij}c_{ij}\phi_i(0)\Big(\phi_m(0)\delta_{j,n} +A^{(2)}_{mn,j}\Big)+D\sum_{ij}c_{ij}\Big(B_{m,i}\delta_{j,n} +B^{+\ (2)}_{mn,ij}+B^{-\ (2)}_{mn,ij}\Big), \label{fermi_eigen_equation_b}
	\end{eqnarray}
where 
\begin{eqnarray}
A^{(2)}_{mn,j}&=&\int d\rho \left(\phi_m(\sqrt{3}\rho/2)+\phi_m(-\sqrt{3}\rho/2)\right)\phi_n(-\rho/2)\phi_j(\rho); \\
B_{m,i}&=&\int_{|r|>r_c} dr \phi_m(r)\phi_i(r)/|r|^3;\\
B^{+\ (2)}_{mn,ij}&=&\int_{|r_+|>r_c} drd\rho \phi_m(r)\phi_n(\rho)\phi_i(r_+)\phi_j(\rho_+)/|r_{+}|^{3};\\
B^{-\ (2)}_{mn,ij}&=&\int_{|r_-|>r_c} drd\rho \phi_m(r)\phi_n(\rho)\phi_i(r_-)\phi_j(\rho_-)/|r_{-}|^{3}.
\end{eqnarray}
Note that a non-zero summation $B^{+\ (2)}_{mn,ij}+B^{-\ (2)}_{mn,ij}$ requires $m$ to be even, and thus the full symmetry of $f(r,\rho)$ based on Eq.~(\ref{f_b}) can be guaranteed. 

For the resulted wavefunction $\Psi(r,\rho)$, its exchange symmetry (Eq.~(\ref{sym})) and boundary condition (same as Eq.~(\ref{BC})) can both be proved similar to the fermion case, which will be neglected here.

\subsection{(c) Numerical details} \label{sec_error}

Due to heavy numerics involved, in our practical calculations we have taken finite cutoffs for both $n$ (level index of $\rho$-motion) and $m$ (level index of $r$-motion).  Specifically, we take $n_{cut}=60$ and $m_{cut}=130$. 
The convergency of $n$ is confirmed by comparing the energy spectra with $n_{cut}=60$ and $80$, which result in negligible energy difference $\sim 10^{-5}\omega$. To check the convergency of $m$, we consider the fermion case without dipolar interaction ($D=0$), where the $m$-degree can be effectively eliminated via Eq.~(\ref{dn}) and the energy $E$ can be obtained through\cite{Amico,Blume}
\begin{equation}
G_nd_n=-\sum_j d_j B_{nj}, 
 \label{m_infty}
\end{equation}
with
\begin{eqnarray}
G_n&=&\frac{1}{g}-\sum_{m=0}^{\infty} \frac{|\phi_m(0)|^2}{E-\epsilon_m-\epsilon_n} =  \frac{1}{g}+\frac{\Gamma(-\nu_n)}{2\Gamma(-\nu_n+1/2)\omega l};\\
B_{nj}&=&\sum_{m=0}^{\infty} \frac{\phi_m(0)A^{(1)}_{mn,j}}{E-\epsilon_m-\epsilon_n}=\int d\rho\ \Phi_d(E-\epsilon_n, \sqrt{3}\rho/2)\phi_n(-\rho/2)\phi_j(\rho).
\end{eqnarray}
Here $\nu_n=E/(2\omega)-n/2-1/4$. We have compared the energy spectra obtained from Eq.~(\ref{m_infty}) with those from Eq.~(\ref{fermi_eigen_equation}) at a finite $m_{cut}=130$ (here $n_{cut}=60$ for both cases). It is found that the energy deviations for sTG branch are around $0.05\omega$, while for the atom-dimer branches are reasonably larger $\sim 0.2\omega$ within the interaction range $\omega l/g<-0.25$. However, the energy gap ($E_G$) at the avoided level crossing of two branches is much less affected by finite $m_{cut}$, as shown by the following table.
\begin{table}[h]
		\centering
		\begin{tabular}{c|c|c|c} 
			\quad&\textbf{n=1} & \textbf{2} & \textbf{3}\\
			\hline
			$E_G^{(m_{cut}=130)}$&5.523 & 1.889 & 0.689\\
			\hline
			$E_G^{(m_{cut}=\infty)}$&5.529 & 1.856 & 0.677\\
		\end{tabular}
		\caption{Energy gap $E_G$ (in unit of $10^{-3}\omega$) obtained from finite and infinite $m_{cut}$ in the case $D=0$. }
		\label{gap_comparison}
\end{table}
\par One can see that the relative errors of $E_G$ are less than $2\%$. 

Moreover, we note that the finite $m_{cut}$  has little effect to the shift $\Delta(1/g_c)$ (Eq.~(7) in the main text). This is because $\partial E_d/\partial (1/g)$ in the denominator is inversely proportional to $\sum_m |\phi_m(0)|^2/(E-\epsilon_m)^2$, which has a fast convergence at large $m$. We have checked that the relative derivations of $\sum_m |\phi_m(0)|^2/(E-\epsilon_m)^2$ between $m_{cut}=130$ and exact results are below $7\%$ for the interaction regime $\omega l/g <-0.25$.

For a finite $D\neq 0$, one cannot eliminate the $m$-degree as in Eq.~(\ref{m_infty}) and thus cannot check its convergence as above. Nevertheless, since we have considered a rather weak $D(\ll \omega l^3)$ in our work, we do not expect the convergency to change too much as compared to $D=0$ case.

For three identical bosons, we have also checked the finite $m_{cut}$ to energy solutions in the case $D=0$. In this case, the three-body equation after eliminating $m$-degree is the same as Eq.~(\ref{m_infty}) except that the coefficient on its right side has to change from $-1$ to $2$. We have compared the energy spectra  obtained from this equation ($m_{cut}=\infty$) with those from a finite $m_{cut}=130$. The energy deviations for sTG branch are around $0.1\omega$, while for the bound state branches are $\sim 0.5\omega$ within the interaction range $\omega l/g<-0.3$. These deviations are all larger than fermion case.
Meanwhile, we find that the energy gap $E_G$ at the avoided level crossing of two branches is little affected by finite $m_{cut}$. For $n=1$, the energy gap $E_G$  is $0.00392\omega$ calculated from finite $m_{cut}=130$ and is $0.00407\omega$ from the exact result. The relative errors of $E_G$ are less than $4\%$.

\begin{figure}[h]
\includegraphics[width=14cm]{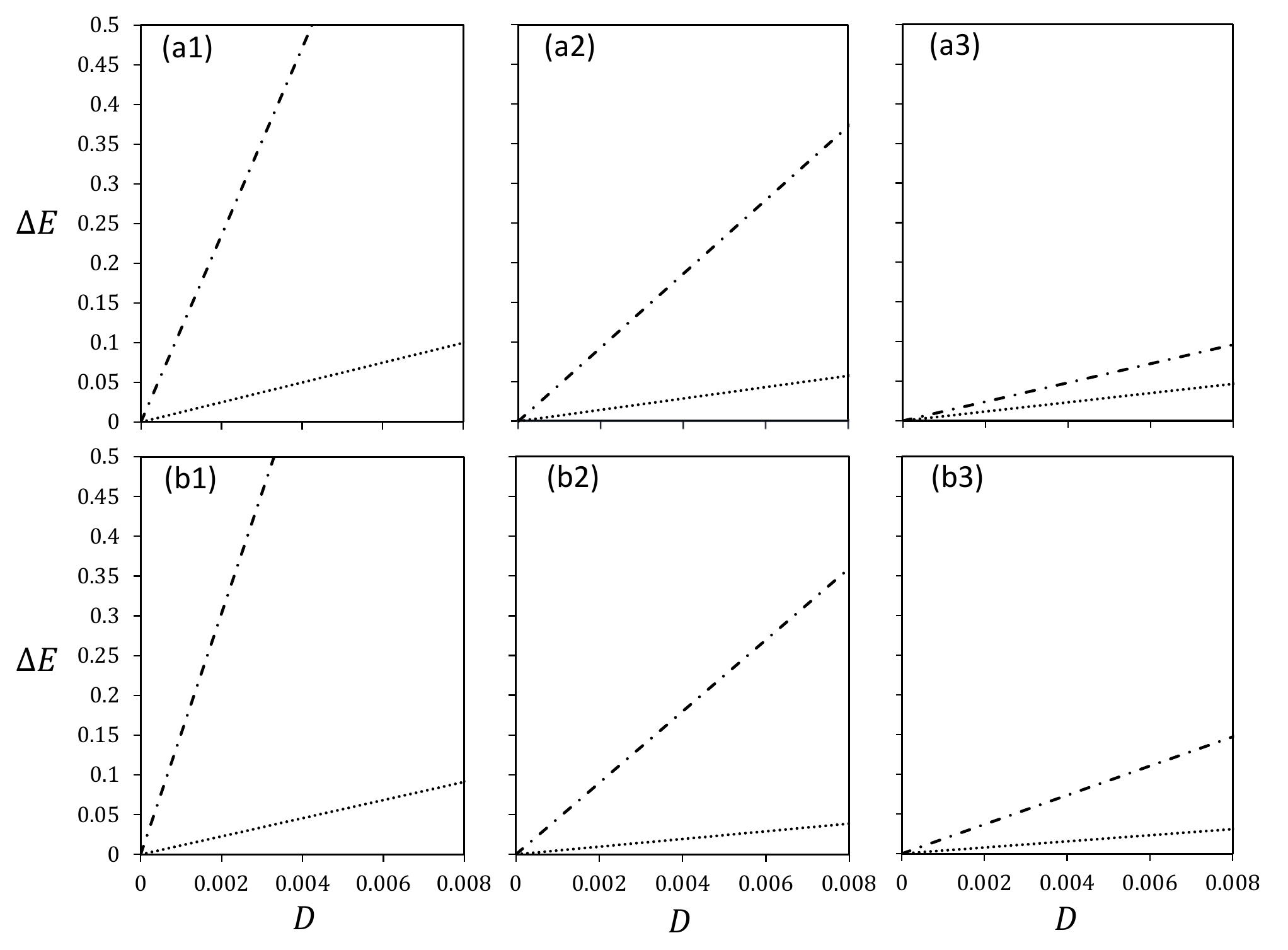}
\caption{(Color online). Energy response $\Delta E\equiv \langle V_{dd}\rangle$ to weak dipolar repulsion ($D>0$) for the sTG (dotted) and excited bound state (dash-dotted) branches of three-fermion (a1-a3) and three-boson (b1-b3) systems at given $\omega l/g=-0.27$. Here we choose different dipolar cutoffs $r_c/l =0.1,\ 0.15,\ 0.2$ in (a1,b1), (a2,b2) and (a3,b3). $\Delta E$,   $g$ and $D$ are respectively in units of $\omega$, $\omega l$ and $\omega l^3$.} 
\label{fig_rc}
\end{figure}

Finally, we discuss the effect of $r_c$, the short-range cutoff of dipolar interaction $V_{dd}$. Here we have neglected the short-range details of the reduced $V_{dd}$ in a realistic quasi-1D system and simplified it as  $V_{dd}= D/|r|^3$ for $r>r_c$ and $=0$ otherwise. In the main text, we have taken $r_c=0.15 l$. In Fig.~\ref{fig_rc}, we further show the energy responses $\Delta E\equiv \langle V_{dd}\rangle$ as functions of $D$ for both sTG and bound state branches of three fermions (a1-a3) and bosons (b1-b3) at different $r_c$. One can see that the fermion and boson systems show similar responses to $r_c$: in comparison to  $r_c=0.15 l$ case (Fig.~\ref{fig_rc}(a2,b2)), a smaller $r_c$ induces a higher $\Delta E$ for both branches (Fig.~\ref{fig_rc}(a1,b1)), and a larger $r_c$ gives a lower $\Delta E$ (Fig.~\ref{fig_rc}(a3,b3)). Moreover, we note that their energy difference is also enhanced as $r_c$ becomes smaller. It is because a smaller $r_c$ can accumulate more  interaction energy for the bound state branch than the sTG branch, given the former is more localized at short range. Therefore the shift of $1/g_c$ (c.f. Eq.~(7) in the main text) should be more dramatic for smaller $r_c$ at a given $D$.

\section{II.   \ \ \ Analytical results for fermionic sTG gas near resonance}

It is convenient to work in the original coordinate space $\{x_1,x_2,x_3\}$ to analyze the fermionic sTG gas near resonance (here $x_1$ is associated with spin $\downarrow$, and $x_2,x_3$ are with spin $\uparrow$). It is known that this system is well described by the effective spin chain Hamiltonian\cite{Zinner, Santos, Pu, Cui2, Parish}  $H=J/g \sum_i ({\bf s}_i\cdot {\bf s}_{i+1}-1/4)$, with $J$ given by
\begin{equation}
J=\frac{12}{m^2}\int dx_1dx_2dx_3 \delta(x_1-x_2)\theta(x_1=x_2<x_3)\left|\frac{\partial \Psi_F}{\partial (x_2-x_1)}\right|^2 .
\end{equation}
Here $\Psi_F=\frac{1}{\sqrt{6}} {\rm Det} (\psi_i (x_j))$ is the ground state wavefunction of non-interacting fermions, and its energy is $E_0=9\omega/2$; $\theta(...)$ is a dimensionless function, and to be specific we have $\theta(x_1=x_2<x_3)=1$ if $x_1=x_2<x_3$ is satisfied and $=0$ otherwise. The spin chain model determines the spin order in spatial space and gives the zero-th order wavefunction as
\begin{eqnarray}
    	\Psi_0(x_1,x_2,x_3)&=&\Psi_F(x_1,x_2,x_3)\times\frac{1}{\sqrt{2}}\Big((\theta(x_1<x_2<x_3)+\theta(x_1<x_3<x_2)-2\theta(x_2<x_1<x_3) \Big. \nonumber\\ 
	&&\Big.-2\theta(x_3<x_1<x_2)+\theta(x_2<x_3<x_1)+\theta(x_3<x_2<x_1)\Big).
    \end{eqnarray}
 The first-order correction ($=-\Psi_1/g$) can be obtained by treating the interaction as zero-th order Hamiltonian and the kinetic term as perturbation\cite{Pu}, which gives    
    \begin{eqnarray}
    	\Psi_1(x_1,x_2,x_3)&=&-\frac{3}{m\sqrt{2}}\frac{\partial \Psi_F}{\partial(x_2-x_1)}(\theta(x_3>x_2=x_1)-\theta(x_3<x_2=x_1))\nonumber\\
    &&-\frac{3}{m\sqrt{2}}\frac{\partial\Psi_F}{\partial(x_3-x_1)}(\theta(x_2>x_3=x_1)-\theta(x_2<x_3=x_1)).
    \end{eqnarray}
This first-order perturbation of wavefunction  leads to the second-order perturbation of energy, as shown in  Eq.~(6) in the main text, which can also be obtained by directly diagonalizing the effective spin chain model.

Up to the leading order of $1/g$, the wavefunction and energy of sTG branch are then $\Psi_{\rm sTG}$ and $E_{sTG}$ as shown in Eqs.~(5,6) in the main text. To facilitate the comparison of sTG branch with other states in the center-of-mass (CoM) frame, we have to transform both $\Psi_{\rm sTG}$ and $E_{sTG}$ into the same frame, which we will denote as $\tilde{\Psi}_{\rm sTG}$ and $\tilde{E}_{sTG}$ respectively. Specifically, we have 
 \begin{equation}
 \tilde{\Psi}_{\rm sTG}(r,\rho)=\Psi_{\rm sTG}/\Phi_{0}(R);\ \ \ \ \tilde{E}_{sTG}={E}_{sTG}-\omega/2,
 \end{equation}
 where $R=(x_1+x_2+x_3)/3$ is the CoM coordinate, and $\Phi_{0}(R)=\pi^{-1/4}l_R^{-1/2}e^{-R^2/(2l_R^2)}H_0(R/l_R)$ represents the ground state of CoM motion (with $l_R=(M\omega)^{-1/2}$ and $M=3m$). In Fig.~2(d3) of the main text, we have plotted the normalized  $\tilde{\Psi}_{\rm sTG}$ for a typical sTG state. There, to improve its visibility we have approximated $\theta(x_i=x_j)$ as a Gaussian function $e^{-(x_i-x_j)^2/\sigma^2}$ with a narrow width $\sigma(\ll l)$. 
    
\section{III. \ \ \ Results of three identical bosons}

\begin{figure}[h]
\includegraphics[width=14cm]{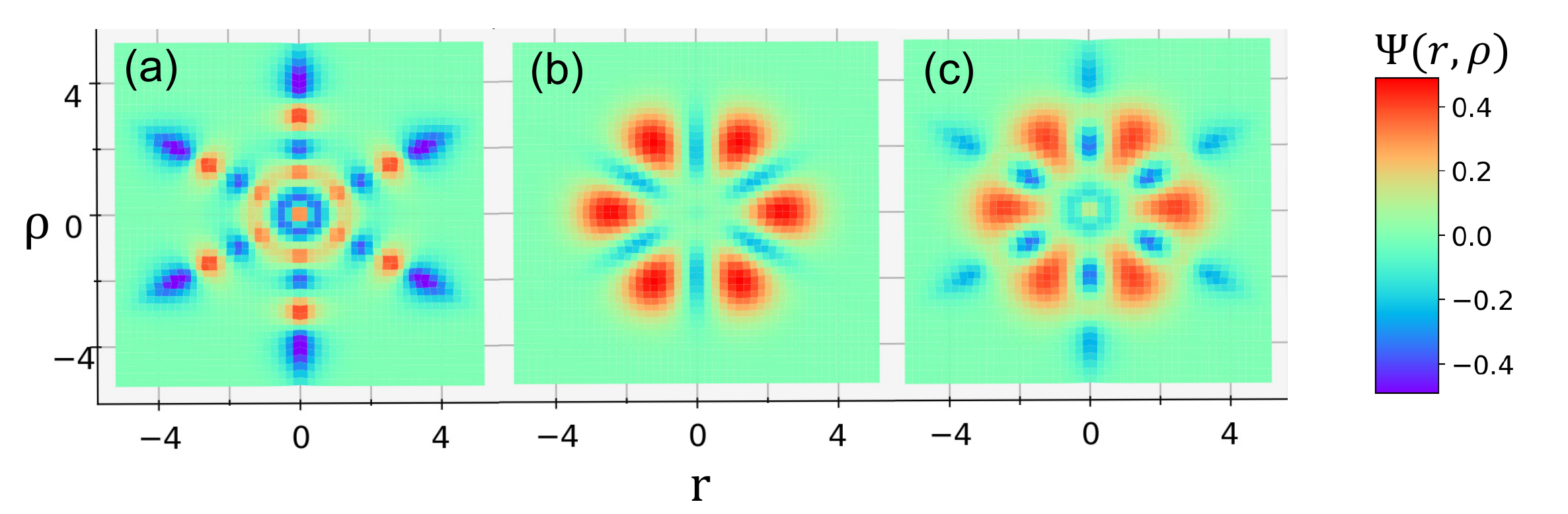}
\caption{(Color online). Contour plots of wave-function $\Psi(r,\rho)$ for three identical bosons near and far from avoided level crossings. (a) and (b) are, respectively, for an excited bound state and the sTG branch at $\omega l/g=-0.35$, away from any level crossing. (c) is for a strongly hybridized state near an avoided level crossing at $\omega l/g=-0.33$. Here we have taken $l$ as the length unit and $l^{-1}$ as the unit of wavefunction  $\Psi$. 
} \label{fig1_supple}
\end{figure}

Here we present our numerical results for three identical bosons. Fig.~\ref{fig1_supple} shows the real-space distribution of $\Psi(r,\rho)$ near and far from the inter-branch (avoided) crossings. The main difference from the fermion case (Fig.~2(c1-c3) in the main text) is that bosons hold a cluster bound state where any pair of three atoms can come close to each other, and thus its wavefunction is peaked at both $r\rightarrow 0$ and $r_{\pm}=(\pm r+\sqrt{3}\rho)/2\rightarrow 0$, see Fig.~\ref{fig1_supple}(a). In this case, the energies of all cluster bound states cannot be estimated using the atom-dimer decomposition as in the fermion case. Moreover, we note that $\Psi(r,\rho)$ in Fig.~\ref{fig1_supple} shows very high symmetries due to Bose statistics. Despite all these differences, there are still important common features between bosons and fermions. Specifically, they all support spatially localized excited bound states and an extended sTG gas (see Fig.~\ref{fig1_supple}(a,b)), and the two states can strongly hybridize with each other near their level crossing (Fig.~\ref{fig1_supple}(c)). This suggests that the loss mechanism of sTG gas in bosons is identical to that in fermions,  which has been elaborated on in the main text.   

\begin{figure}[t]
\includegraphics[width=13cm]{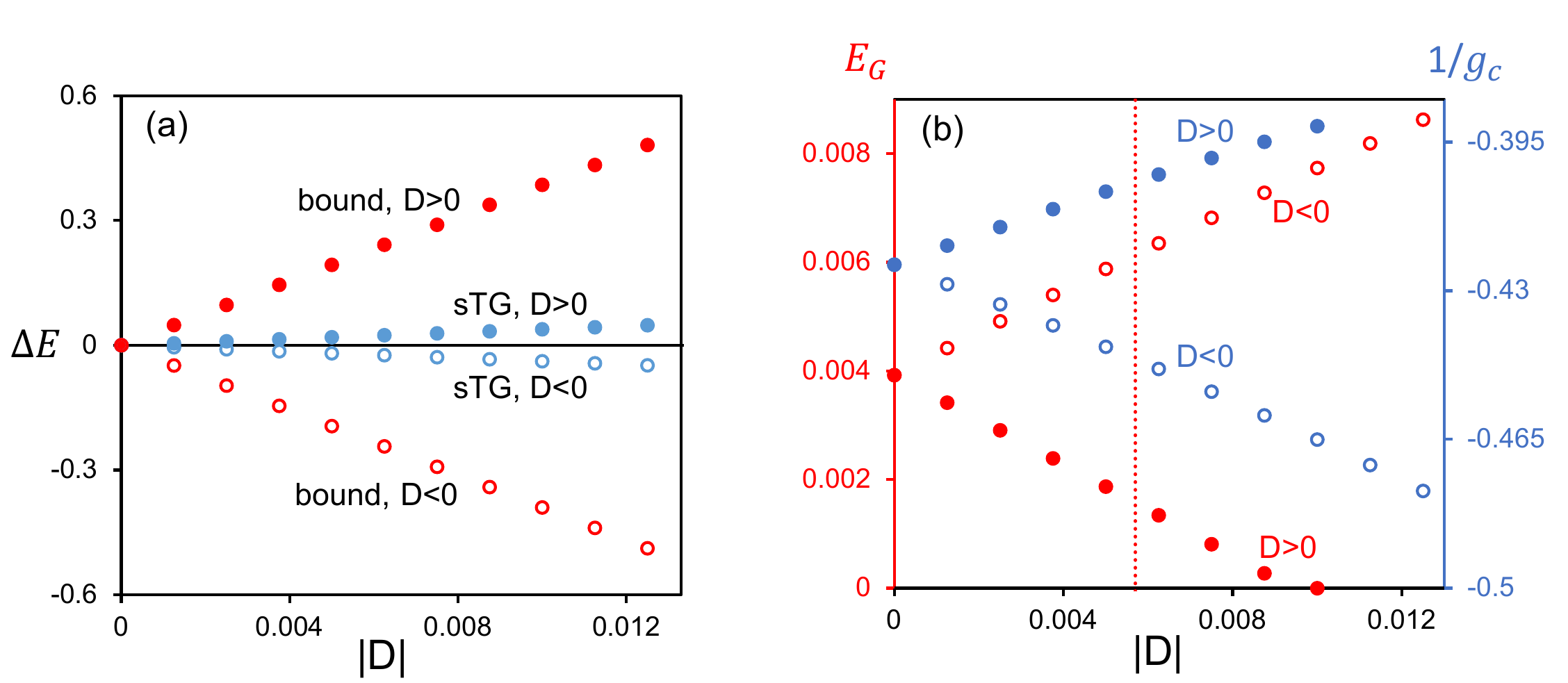}
\caption{(Color online). Response of three identical bosons to a weak dipolar interaction with strength $D$. (a) Energy shifts of sTG gas and its nearby bound state branch as functions of $|D|$ at given $\omega l/g=-0.3$. (b) Energy gap and location of the first avoided crossing as functions of $|D|$.  Dotted vertical line marks the strength of repulsive $D$ used in experiment.
Here the energy $E$,   coupling $g$, and dipolar force $D$ are respectively in units of $\omega$, $\omega l$ and $\omega l^3$.} \label{fig2_supple}
\end{figure}

To further check our expectation,  in Fig.~\ref{fig2_supple}(a) we take a typical $g$ away from any level crossing and plot out the energy shifts $\Delta E$ for the sTG gas and its nearest bound state as varying $|D|$. We can see that these two branches behave very differently in the presence of dipolar interaction. Namely, the bound state energy changes much more rapidly than the sTG energy when increasing $|D|$. This is consistent with fermion case, and again can be attributed to distinct real-space distributions of the two branches (Fig.~\ref{fig1_supple}). 
  
In Fig.~\ref{fig2_supple}(b) we show the evolutions of energy gap ($E_G$) and location ($1/g_c$)  for the first avoided crossing (furthest from resonance) in the spectrum as $|D|$ increases.  Again consistent with the fermion case, for a repulsive $D>0$, $1/g_c$ shifts to resonance side with a decreasing $E_G$, while for an attractive $D<0$,  $1/g_c$ shifts away from resonance with an increasing $E_G$. This indicates a more (less) stable sTG gas when adding a repulsive (attractive) dipolar interaction. Taking the same value of repulsive $D$ as in the experiment, as marked by vertical dotted line in Fig.~\ref{fig2_supple}(b), we note that $E_G$ can be reduced to nearly one-third of its original value at $D=0$, similar to the fermion case.

\end{document}